\documentclass[10pt,prl,twocolumn,superscriptaddress,byrevtex,showpacs]{revtex4-1}
\usepackage{graphics}
\usepackage{amsmath}
\pdfoutput=1

\newcommand{\be}{\begin{equation}}
\newcommand{\ee}{\end{equation}}
\newcommand{\om}{\omega}
\newcommand{\vp}{\varphi}
\newcommand{\ra}{\rightarrow}
\newcommand{\D}{\mathrm{d}}
\newcommand{\I}{\mathrm{i}}
\newcommand{\E}{\mathrm{e}}
\newcommand{\lpole}{\times}
\newcommand{\lsp}{*}
\newcommand{\B}{\mathcal{B}}

\begin{document}

\title{First-order phase transitions from poles in asymptotic\\ representations of partition functions}

\author{Hugo Touchette}
\email{ht@maths.qmul.ac.uk}
\author{Rosemary J. Harris}
\address{School of Mathematical Sciences, Queen Mary University of London, London E1 4NS, UK}

\author{Julien Tailleur}
\address{School of Physics and Astronomy, University of Edinburgh, Edinburgh EH9 3JZ, UK}

\date{\today}

\begin{abstract}
Although partition functions of finite-size systems are always analytic, and hence have no poles, they can be expressed in many cases as series containing terms with poles. Here we show that such poles can be related to linear branches of the entropy, expressed in the thermodynamic limit as a function of the energy per particle. We also show that these poles can be used to determine whether the entropy is nonconcave or has linear parts, which is something that cannot be done with the sole knowledge of the thermodynamic free energy derived from the partition function. We discuss applications for equilibrium systems having first-order phase transitions.
\end{abstract}


\maketitle

It is known from the seminal work of Lee and Yang \cite{yang1952} that the zeros of partition functions, seen as functions of a complex variable, provide useful information about the analyticity properties of the free energy in the thermodynamic limit, and thus about the appearance of phase transitions in that limit. Their main results, extended by Fisher \cite{fisher1965} to the canonical partition function,
\be
Z(\beta)=\sum_{\text{microstates}\ \om} \E^{-\beta H(\om)},
\ee
of an $N$-particle system with Hamiltonian $H$, show that the complex zeros of $Z(\beta)$ lie away from the real axis for all $N<\infty$, and that, in the presence of a phase transition, the zeros of $Z(\beta)$ get infinitesimally close to the real axis in the limit $N\ra\infty$. The real value $\beta_c$ at which the zeros ``pinch'' the real axis in this limit corresponds to the critical inverse temperature at which the thermodynamic free energy function, defined as
\be
\vp(\beta)=\lim_{N\ra\infty}-\frac{1}{N}\ln Z(\beta),
\label{eqf1}
\ee
is non-analytic \footnote{For practical reasons, we define the free energy without the additional factor $1/\beta$.}. Moreover, the angle at which the locus of zeros crosses the real axis determines the order of the phase transition corresponding to $\beta_c$ \footnote{Similar results hold for the grand-canonical ensemble, which is the original ensemble used by Lee and Yang \cite{yang1952}.}. These properties of the zeros of $Z(\beta)$ have been widely used for studying phase transitions in equilibrium systems (see \cite{bena2005} for a review), as well as, more recently, nonequilibrium systems in steady states \cite{arndt2000,*blythe2002}.

We study in this paper another component of $Z(\beta)$, which we refer to as the set of ``poles'' of $Z(\beta)$, and relate it to the thermodynamic properties of the system represented by $H$. We use the term ``pole'' with quotes because $Z(\beta)$ has of course no singular points, since it is an analytic function of $\beta$ for $N$ finite. The poles that we have in mind actually appear in the terms of a special asymptotic expansion of $Z(\beta)$ to be defined below. The main result that we prove here is that these poles are related to linear branches of the microcanonical entropy of the system represented by $H$, and can therefore be used to study two physical phenomena associated with these branches, namely, first-order phase transitions and phase separation in the canonical ensemble \cite{touchette2009}.

The knowledge of these poles also solves an outstanding problem in the field of long-range interacting systems \cite{draw2002,*campa2009}, which is to determine whether a system undergoing a first-order phase transition in the canonical ensemble has an entropy in the microcanonical ensemble which is nonconcave or is concave but has a linear branch. These two types of entropy are known to lead to the same non-differentiable free energy function $\varphi(\beta)$ \cite{touchette2009}, so they cannot be distinguished from the point of view of this function alone. Our results show, however, that they can be distinguished using information about the ``poles'' of $Z(\beta)$. As an illustration of these results, we compute the linear entropy of a simple model of DNA denaturation from its partition function. At the end, we also comment on the applicability of our results for calculating large deviation functions characterizing the fluctuations of nonequilibrium systems.

The problem that we are concerned with is to calculate the microcanonical entropy function
\be
s(u)=\lim_{N\ra\infty}\frac{1}{N}\ln \Omega(u)
\ee
from the density of states $\Omega(u)$, which is obtained from $Z(\beta)$ via the inverse Laplace transform formula
\be
\Omega(u)=\frac{1}{2\pi \I}\int_{r-\I\infty}^{r+\I\infty} Z(\beta)\, \E^{\beta Nu}\, \D\beta,
\label{eqlt1}
\ee
where $r$ is an arbitrary real number located inside the region of convergence of $Z(\beta)$ \cite{kubo1965}. Given that $s(u)$ is a thermodynamic-limit quantity, it is often assumed that its calculation via Eq.~(\ref{eqlt1}) requires not the exact knowledge of $Z(\beta)$, but only of the asymptotic behavior of $Z(\beta)$ as $N\ra\infty$, expressed, according to Eq.~(\ref{eqf1}), as 
$
Z(\beta)\approx \E^{-N\vp(\beta)}
$
with sub-exponential corrections in $N$. By substituting this asymptotic result in the inverse Laplace transform, and by performing a saddlepoint approximation of the complex integral, one indeed finds
\be
s(u)=\inf_\beta \{\beta u-\vp(\beta)\}.
\ee

The problem with this result, which is nothing but a Legendre transform written in a technical form, is that it does not always hold \cite{touchette2009}. To see why, consider the following two ``mock'' partition functions:
\be
Z_{1}(\beta)=\E^{N\beta}+\E^{-N\beta},\quad Z_{2}(\beta)=\frac{\E^{N\beta}-\E^{-N\beta}}{\beta}.
\label{eqz1}
\ee
It is easily verified that each of these partition functions is analytic, despite the appearance of the $1/\beta$ term in the second, and that both partition functions lead to the same free energy function $\vp(\beta)=-|\beta|$. However, the densities of states obtained from $Z_{1}(\beta)$ and $Z_{2}(\beta)$ and their corresponding entropies must be different, since the partition functions are themselves different. A simple calculation of the inverse Laplace transform shows that this is the case. The density of states obtained from $Z_{1}(\beta)$ is $\Omega_{1}(u)=\delta(u+1)+\delta(u-1)$, whereas the density of states obtained from $Z_{2}(\beta)$ is such that $\Omega_{2}(u)=1$ for $u\in [-1,1]$ and is $0$ otherwise. As a result, the entropy $s_1(u)$ obtained from $\Omega_{1}(u)$ is finite and zero only for $u=\pm 1$, whereas the entropy $s_2(u)$ associated with $\Omega_{2}(u)$ is finite and zero for all $u\in [-1,1]$. Hence the entropies calculated from $Z_{1}(\beta)$ and $Z_{2}(\beta)$ are different, but both partition functions lead to the same free energy $\vp(\beta)$.

The problem illustrated by this example is well documented in large deviation theory \cite{touchette2009}, and has been discussed recently in the context of long-range interaction systems, such as self-gravitating particles and unscreened plasmas, as these often have nonconcave entropies \cite{draw2002,*campa2009}. The problem is that entropy functions that have the same concave envelope, as in the example above, have the same free energy $\vp(\beta)$, and cannot, as mentioned before, be distinguished from the point of view of $\vp(\beta)$ alone. This means concretely that the knowledge of the asymptotic behavior $Z(\beta)\approx\E^{-N\varphi(\beta)}$ is not sufficient in general to compute $s(u)$; extra information is required to faithfully obtain $s(u)$, especially if one suspects that $s(u)$ is nonconcave or has a linear branch.

The two partition functions shown in Eq.~(\ref{eqz1}) give a hint as to what kind of extra information is required. By recasting each of these partition functions in the form
\be
Z(\beta)=a(\beta)\, \E^{N\beta}+b(\beta)\, \E^{-N\beta},
\label{eqans1}
\ee
we see that what distinguishes $Z_{1}(\beta)$ from $Z_{2}(\beta)$ is the presence of poles in the coefficients $a(\beta)$ and $b(\beta)$ of $Z_{2}(\beta)$. As we show next, it is the presence of these poles in the series representation of $Z_{2}(\beta)$ that is responsible for the linear behavior of $s_2(u)$ seen for $u\in[-1,1]$. This applies to any partition function, in the sense that poles in asymptotic expansions of $Z(\beta)$ are generally associated with linear branches of $s(u)$. 

To demonstrate this claim, we go back to the formula of the inverse Laplace transform shown in Eq.~(\ref{eqlt1}), and assume, as an extension of Eq.~(\ref{eqans1}), that $Z(\beta)$ admits an asymptotic expansion of the form
\be
Z(\beta)= \sum_{j} c_{j}(\beta)\, \E^{-N\vp_j(\beta)}.
\label{eqexp1}
\ee
This expansion can always be obtained for 1D systems by expanding $Z(\beta)$, for example, in the eigenbasis of the transfer matrix associated with $H$ \cite{baxter1982}. For systems of higher dimensions, there is not necessarily a transfer matrix to work with, and for these, Eq.~(\ref{eqexp1}) should presently be considered as an ansatz rather than a derived result. This point will be discussed in more detail in a subsequent paper \footnote{H. Touchette, R. J. Harris, J. Tailleur, in preparation (2010).}. 

Here we shall work on the assumption that Eq.~(\ref{eqexp1}) is given, and that the functions $\vp_j(\beta)$ obtained are concave and smooth functions of $\beta$ that do not depend on $N$. Moreover, we shall assume that the coefficients $c_{j}(\beta)$ are sub-exponential in $N$, and may have poles in the complex $\beta$-plane \footnote{The fact that the coefficients $c_j(\beta)$ are allowed to have poles does not mean, of course, that $Z(\beta)$ has poles; see, e.g., $Z_{2}(\beta)$ in Eq.~(\ref{eqz1}). Note also that the expansion shown in Eq.~(\ref{eqexp1}) is not unique, as poles can be introduced simply by multiplying $Z(\beta)$ by $(\beta-a)/(\beta-a)$, and by distributing the numerator of the resulting expression. Obviously, the addition of such poles does not change $Z(\beta)$.}. These assumptions are verified for some models of interest [10], including the one studied at the end of this paper.

It should be mentioned that expansions similar to Eq.~(\ref{eqexp1}) have been considered before in studies of Yang-Lee zeros, first-order phase transitions, and metastability (see, e.g., \cite{borgs1992,*borgs1992a,*biskup2000}). However, to our knowledge, none have considered the possibility that the coefficients $c_j(\beta)$ may have poles in $\beta$. To see how these poles relate to the properties of $\Omega(u)$ and in turn $s(u)$, we insert the expansion (\ref{eqexp1}) in the formula of the inverse Laplace equation, and proceed to evaluate the complex integral by going through the following steps:

1.\ Distribute the integral of the inverse Laplace transform inside the sum of the partition function to obtain
\be
\Omega(u)=\sum_j\frac{1}{2\pi \I}\int_{r-\I\infty}^{r+\I\infty} c_{j}(\beta)\, \E^{N[\beta u-\vp_j(\beta)]}\,\D\beta.
\label{eqint2}
\ee
This is permitted provided that all the integrals inside the sum converge. With this in mind, we should choose $r$ in such a way that each integral in Eq.~(\ref{eqint2}) is convergent. In particular, we cannot put $r$ on any poles of $c_{j}(\beta)$.

2.\ Approximate each of the integrals labeled by $j$ in Eq.~(\ref{eqint2}) to exponential order in $N$ using the saddlepoint or steepest descent approximation \cite{bender1978}. This requires that we deform the vertical integration contour sitting at $r$, which is often called the \emph{Bromwich contour}, to another equivalent steepest-descent contour that passes through the saddlepoint of the exponent 
\be
\Phi_j(\beta,u)=\beta u-\vp_j(\beta),
\ee
in such a way that $\textrm{Im}\,\Phi_j(\beta,u)$ is constant. Assuming that $\vp_j(\beta)$ is differentiable and concave, the saddlepoint is given by the unique solution of $\varphi'_j(\beta)=u$. Henceforth, we denote this saddlepoint by $\beta_j^\lsp$, and the steepest-descent contour passing through this point by $D_j$.

\begin{figure}[t]
\resizebox{3.4in}{!}{\includegraphics{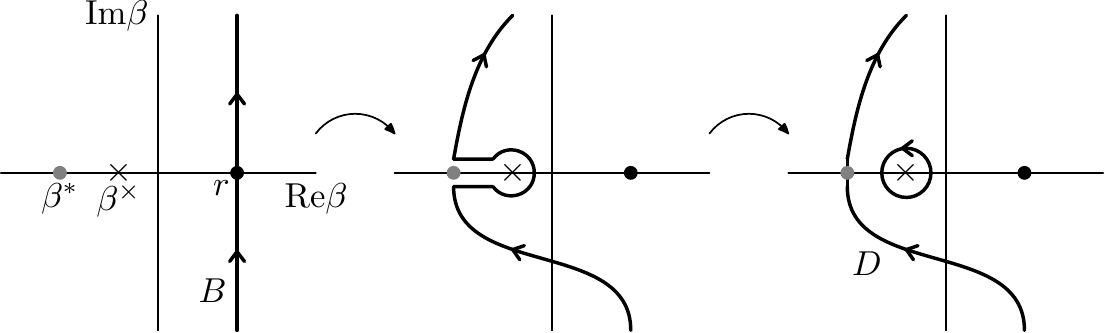}}
\caption{Deformation of the Bromwich contour $B$ sitting at $r$ to a steepest-descent contour $D$ crossing the saddlepoint $\beta^\lsp$. A residue arises if the deformation crosses a pole $\beta^\lpole$.}
\label{figint1}
\end{figure}

3.\ Two situations will arise from the previous step, depending on whether or not the deformation of the Bromwich contour to the steepest-descent contour necessitates that we cross poles of $c_{j}(\beta)$. On the one hand, if no such poles need to be crossed, then the integral on the Bromwich contour is equivalent to the integral evaluated on $D_j$. On the other hand, if the deformation requires that we cross any poles of $c_j(\beta)$, then 
\begin{eqnarray}
\frac{1}{2\pi\I}\int_B c_{j}(\beta)\, \E^{N\Phi_j(\beta, u)}\,\D\beta &=&
\frac{1}{2\pi\I}\int_{D_j} c_{j}(\beta)\, \E^{N\Phi_j(\beta, u)}\,\D\beta\nonumber\\
&& \qquad+\sum\text{Res},
\label{eqint4}
\end{eqnarray}
where $\sum\text{Res}$ is the sum of the residues of the poles that were crossed when transforming the Bromwich contour $B$ into the steepest-descent contour $D_j$ (see Fig.~\ref{figint1}). 

4.\ Insert the result obtained in (\ref{eqint4}) into the sum of Eq.~(\ref{eqint2}), and cancel any terms that have the same magnitude but opposite sign. At this point, we expect many residue terms to cancel. What remains can be put in the form
\be
\Omega(u)=\sum_j\left( \frac{1}{2\pi\I}\int_{D_j} c_{j}(\beta)\, \E^{N\Phi_j(\beta, u)}\,\D\beta+\sum_{\ell} \text{Res}(\beta_{j\ell}^\lpole)\right),
\ee
where $\text{Res}(\beta_{j\ell}^\lpole)$ denotes the residue of $c_j(\beta)\,\E^{N\Phi_j(\beta,u)}$ for the pole $\beta_{j\ell}^\lpole$.

5.\ Approximate the integrals on the steepest-descent contours $D_j$ by their saddlepoints, i.e.,
\be
\frac{1}{2\pi\I}\int_{D_j} c_{j}(\beta)\, \E^{N\Phi_j(\beta,u)}\,\D\beta\approx \E^{N\Phi_j(\beta_j^\lsp,u)}
\label{eqint3}
\ee
with sub-exponential corrections in $N$ \cite{bender1978}. The term $c_{j}(\beta)$ does not contribute to the approximation because it is assumed to be sub-exponential with $N$.

6.\ Evaluate the residue terms. If we assume, for simplicity, that the $c_{j}(\beta)$'s have only simple poles, then the residues are approximately given by
\be
\text{Res}(\beta_{j\ell}^\lpole)\approx\sigma_{j\ell}\,\E^{N\Phi_j(\beta^\lpole_{j\ell}, u)},
\ee
with sub-exponential corrections in $N$, where $\sigma_{j\ell}=\pm 1$. Note that this approximation must be performed only for those poles $\beta^\lpole_{j\ell}$ that were crossed in Step 3 and do not get cancelled in Step 4 above. The remarkable feature of these poles is that they give rise to terms that are exponential in $N$ similar to the saddlepoints.

7.\ The result of Steps 5 and 6 is the following approximation for the density of states:
\be
\Omega(u)\approx\sum_j \left(\E^{N\Phi_j(\beta_j^\lsp, u)}+\sum_\ell \sigma_{j\ell}\,\E^{N\Phi_j(\beta^\lpole_{j\ell}, u)}\right).
\label{eqds1}
\ee
As a last step, we further approximate this expression by taking the largest exponential term (Laplace approximation). In order to express this final step in a convenient form, let us define $\B_j$ to be the set containing the saddlepoint $\beta_j^\lsp$ and the poles $\beta^\lpole_{j\ell}$ that remain after Step 4. Then, by taking the largest term in that expression, we obtain
\be
s(u)=\sup_j \sup_{\beta\in \B_j}\{ \beta u-\vp_j(\beta)\}.
\label{eqs1}
\ee
Note that the sign variable $\sigma_{j\ell}$ does not appear in the above result because the dominant term of $\Omega(u)$ is necessarily positive. Moreover, the saddlepoint or pole selected from the maximization over $\B_j$ is necessarily real, for otherwise $s(u)$ would not be a real function.

The representation of the entropy $s(u)$ shown in Eq.~(\ref{eqs1}) is the main result of this paper. The concavity properties of $s(u)$ are deduced from this equation by analyzing the maximization over the set $\B_j$. Three cases must be distinguished:

\emph{Case 1}:\ If, for an interval $\mathcal{U}$ of values for $u$, the maximization over $\B_j$ picks up a pole, then $s(u)$ will be proportional to $u$ over $\mathcal{U}$ (since poles of $c_j(\beta)$ do not depend on $u$). In this case, $s(u)$ will thus have a linear branch.

\emph{Case 2}:\ If, for $u\in\mathcal{U}$, the maximization over $\B_j$ does not pick up a pole, but picks up instead a saddlepoint $\beta_j^\lsp$ which is constant as a function of $u$, then $s(u)$ will also have a linear branch over $\mathcal{U}$.

\emph{Case 3}:\ If, for $u\in\mathcal{U}$, the maximization over $\B_j$ picks up neither a pole (Case 1) nor a saddlepoint $\beta_j^\lsp$ which is constant in $u$ (Case 2), then $s(u)$ will vary in a nonlinear way with $u$. In this case, $s(u)$ may be concave or nonconcave, but has no linear branch over $\mathcal{U}$.

It can be checked that the linear entropy $s_2(u)$ mentioned earlier arises from Case 1 above. The pole in the corresponding partition function $Z_{2}(\beta)$ is at $\beta=0$ and gives rise to the flat part of $s(u)$ with slope 0. Case 3 applies, on the other hand, to $Z_{1}(\beta)$, as the latter has no pole in its series representation. For examples of entropy calculations involving constant saddlepoints, see \cite{farago2002,kastner2009}. 

To provide an explicit illustration of Case 1 involving poles, we now calculate the entropy of a simple model of DNA denaturation due to Kittel~\cite{kittel1969,*gibbs1959,*cuesta2004}. The partition function of this model can be written in the thermodynamic limit as
\be
Z(\beta)=\frac{1}{\beta-\beta_c}-\frac{\E^{-N(\beta\epsilon-\ln G)}}{\beta-\beta_c},
\label{eqkz1}
\ee
where $\epsilon$ is the energy associated with one bond of a DNA chain consisting of $N$ bonds, $G$ is a degeneracy factor associated with this energy, and $\beta_c=\epsilon^{-1}\ln G$ is the critical inverse temperature at which the thermodynamic free energy $\varphi(\beta)$, derived from Eq.~(\ref{eqf1}), is nondifferentiable \footnote{The partition function shown in Eq.~(\ref{eqkz1}) is actually a variant of the exact partition function of Kittel \cite{kittel1969}, obtained by treating the mean energy $u=U/N$ as a continuous variable in the thermodynamic limit.}. From the Yang-Lee theory point of view, $\beta_c$ is also the accumulation point of the zeros of $Z(\beta)$. 

To obtain $s(u)$ for this model, we first note that $Z(\beta)$ in Eq.~(\ref{eqkz1}) has the form of Eq.~(\ref{eqexp1}) with $\varphi_1(\beta)=0$ and $\varphi_2(\beta)=\beta\epsilon-\ln G$, and that each of the two terms composing $Z(\beta)$ has a pole at $\beta_c$. Next we choose $r>\beta_c$, and follow the calculation steps described above. For $u\in [0,\epsilon)$, we find that the Bromwich integral involving the first term of $Z(\beta)$ gives rise to a residue proportional to $\E^{N\beta_c u}$ because of the pole at $\beta_c$, whereas the Bromwich integral of the second term vanishes \footnote{The Bromwich integrals involving $\varphi_1(\beta)$ and $\varphi_2(\beta)$ can be calculated exactly by properly closing the Bromwich contour, so there is no saddlepoint approximation needed here; more details will be given in a separate paper.}. Therefore, $s(u)=\beta_c u$ for $u\in[0,\epsilon)$. For all other values of $u$, the Bromwich integrals either cancel one another or vanish, and so we find $s(u)=-\infty$ outside $[0,\epsilon)$. This agrees with the entropy that one would obtain by combinatorial means. Moreover, the fact that the finite part of $s(u)$ is linear with slope $\beta_c$ confirms the fact that $\varphi(\beta)$ is nondifferentiable at $\beta_c$ \cite{touchette2009}. 

This calculation of $s(u)$, although simple, provides an illustration of what should be observed in more realistic equilibrium systems with first-order transitions, especially systems involving short-range interactions, such as nearest-neighbor spin systems or screened Coulomb systems \cite{draw2002,*campa2009}. For these, it is known that $s(u)$ is in general a concave function in the thermodynamic limit \cite{ruelle1969}. This implies that, if $\varphi(\beta)$ is nondifferentiable, then $s(u)$ will have in general one or more linear branches \cite{touchette2009}, which are likely to be associated, according to our results, with poles in some expansion of $Z(\beta)$.

It should be obvious, to conclude, that our results can be applied to partition functions other than the canonical one to calculate the entropy as a function of macrostates other than the energy per particle (e.g., magnetization or particle density). Our results can also be generalized, following the theory of large deviations \cite{touchette2009}, to calculate entropy functions describing the fluctuations of observables of nonequilibrium systems in driven steady states. In this context, one must replace $Z(\beta)$ by the generating function of the observable considered. Poles in series representations of generating functions have been considered in the context of nonequilibrium systems (see, e.g., \cite{farago2002,zon2003,*harris2006,*rakos2008,*visco2006a}), and are known to be associated with extensions of the Gallavotti-Cohen symmetry of nonequilibrium fluctuations \cite{gallavotti1995}. We expect such poles to also play a role in nonequilibrium first-order phase transitions, as these are generally characterized by nonequilibrium entropy functions (viz., rate functions) having linear or nonconvex branches. 

\begin{acknowledgments}
We thank O. Bandtlow, S. Grosskinsky, M. Kastner, and S. Ruffo for useful discussions. The work of H.T. is supported by an RCUK Academic Fellowship.
\end{acknowledgments}

%\bibliography{touchettepolesprl1}
\bibliography{masterbibmin}

\end{document}